\documentclass{jpp}
\usepackage{graphicx,natbib}

\ifCUPmtlplainloaded \else
  \checkfont{eurm10}
  \iffontfound
    \IfFileExists{upmath.sty}
      {\typeout{^^JFound AMS Euler Roman fonts on the system,
                   using the 'upmath' package.^^J}%
       \usepackage{upmath}}
      {\typeout{^^JFound AMS Euler Roman fonts on the system, but you
                   dont seem to have the}%
       \typeout{'upmath' package installed. JPP.cls can take advantage
                 of these fonts, if you use 'upmath' package.^^J}%
      }
  \else
  \fi
\fi

\usepackage{amsmath,amssymb}

\begin{document}
\title[Proton fire hoses in the solar wind]{Proton fire hose instabilities \\ in the expanding solar wind}

\author{Petr Hellinger}

\affiliation{Astronomical Institute, CAS,
Bocni II/1401, CZ-14000 Prague, Czech Republic\\[\affilskip]
Institute of Atmospheric Physics, CAS,
Bocni II/1401, CZ-14000 Prague, Czech Republic
}

\pubyear{2016}
\volume{?}
\pagerange{??}
\date{??}

\maketitle

\begin{abstract}

Using two-dimensional hybrid expanding box simulations we study
the competition between the continuously driven parallel proton temperature
anisotropy and fire hose instabilities  in collisionless homogeneous plasmas.
For quasi radial ambient magnetic field
the expansion drives $T_{\mathrm{p}\|}>T_{\mathrm{p}\perp}$ and
the system becomes eventually unstable with respect to the dominant parallel fire hose
instability. This instability is generally unable to counteract the induced
anisotropization and the system typically becomes unstable with respect
to the oblique fire hose instability later on. The oblique instability efficiently reduces
the anisotropy and the system rapidly stabilizes while a significant part of the generated
electromagnetic fluctuations is damped to protons. As long as the magnetic
field is in the quasi radial direction, this evolution repeats itself
and the electromagnetic fluctuations accumulate. For sufficiently oblique
magnetic field the expansion drives $T_{\mathrm{p}\perp}>T_{\mathrm{p}\|}$ and brings the system to
the stable region with respect to the fire hose instabilities.
\end{abstract}

\section{Introduction}
\label{intro}

In situ observations in the weakly collisional solar wind \citep{mars06} indicate the presence of apparent 
bounds on different particle parameters. The bounds on the
temperature anisotropy of protons
\citep{garyal01c,kaspal03,hellal06,hetr14}, 
alpha particles
\citep{marual12},
electrons
\citep{stveal08},
and/or on the
differential velocity between ion species
\citep{boural13,mattal15b},
the differential velocity between different electron species (heat flux driven instabilities)
\citep{garyal99}
are often compatible with the constraints imposed by kinetic plasma instabilities.
Some observations also indicate that there are enhanced levels of fluctuations
near the apparent bounds
\citep{wickal13,lacoal14}
supporting the interpretation that these bounds are imposed by the kinetic
instabilities.

The evolution of the solar wind particle properties depend on the processes
that are active. For instance, Alfv\'en ion cyclotron waves heat the ions in the
direction  perpendicular to the ambient magnetic field
 \citep{hois02,hellal05} whereas plasma turbulence generates a range of
particle temperature anisotropies and even nongytropies \citep{serval15}. 
The solar wind expansion tends to generate particle temperature anisotropies
and influences as well the differential streaming between different species.
\cite{mattal07} investigated the radial evolution of the proton temperature
anisotropy in the fast solar wind in the beta-anisotropy plane: 
below 1~au, protons follow an anticorrelation between the proton beta and the 
proton temperature anisotropy
\cite[cf.][]{marsal04}. This behaviour is not compatible with the double adiabatic prediction
\citep{chewal56} and indicates the presence of proton heating, likely owing to the turbulent
cascade 
\citep{cranal09,hellal13}.
Beyond 1 au, the results of \cite{mattal07} indicate that the behaviour of protons changes; the 
system seems to follow in the beta-anisotropy plane a marginal stability path with respect to the proton fire
hose instabilities.

Results similar to the observations of \cite{mattal07} are seen in numerical simulations
using the expanding box model \citep{mattal06,hetr08}: The expansion
drives the parallel temperature anisotropy and fire hose instabilities limit the
accessible temperature anisotropy.
These results were, however, obtained for the strictly radial ambient magnetic field
whereas the magnetic field in the solar wind is generally oblique with respect to
the radial direction and rotates towards a transverse direction \citep{park58}.
For a solar wind magnetic field which is
sufficiently oblique relative to the radial direction from the Sun,
the double adiabatic approximation
 predicts  generation of the perpendicular temperature anisotropy \citep{mattal12}.
In this paper we investigate the role of the oblique magnetic field on the
fire hose instabilities using two-dimensional hybrid expanding box simulations. 
The paper is organized as follows: section~\ref{simul}
presents the simulation model (subsection~\ref{heb}) and the results
for the radial magnetic field for different expansion times (subsections~\ref{radial}
and~\ref{radialslow}). Subsection~\ref{oblique} decribes the results for the Parker spiral
magnetic field. Finally, section~\ref{discussion}
summarizes and discusses the simulation results.

\section{Simulation results}
\label{simul}

\subsection{Expanding box model}

\label{heb}

Here we use 
the expanding box model \citep{grapal93,grve96}
implemented to the
hybrid code developed by \cite{matt94} 
to study the response of a collisionless plasma to a slow expansion.
In this Hybrid Expanding Box (HEB) model
the expansion is described as an external force,
one assumes a solar wind with a
constant radial velocity $v_{sw}$.
Transverse scales (with respect to the radial direction)
 of a small portion of plasma, co-moving with the solar
wind velocity, increase  with time as
$1+t/t_e$, where
$t_e=R_0/v_{sw}$ is the (initial) characteristic expansion time ($R_0$ being
the initial radial distance).
The expanding box uses these co-moving coordinates,
the physical transverse scales of the simulation
box increase with time \cite[see][for a detailed description
of the code]{hetr05} and standard periodic boundary conditions are used.

The characteristic spatial and temporal units used in the model
are the initial proton inertial length $d_{\mathrm{p}0}=c/\omega_{p\mathrm{p}0}$ and
the inverse initial proton cyclotron frequency $1/\omega_{c\mathrm{p}0}$.
Here $c$ is the speed of light, $\omega_{p\mathrm{p}0} = ({n_{\mathrm{p}0}
e^2}/{m_\mathrm{p}\epsilon_0})^{1/2}$ is the initial proton plasma
frequency, $\omega_{c\mathrm{p}0} = {e B_{0}}/{m_\mathrm{p}}$,
$B_{0}$ is the initial magnitude of the ambient magnetic
 field $\boldsymbol{B}_0$,
$n_{\mathrm{p}0}$ is the initial proton density,
$e$ and $m_\mathrm{p}$ are the proton electric
charge and mass, respectively; finally,
$\epsilon_0$ is the dielectric permittivity of
vacuum.
We use a spatial resolution
$\Delta x = \Delta y = 0.5 d_{\mathrm{p}0}$, and there are initially 4,096 particles per cell
for protons.
Fields and moments are defined on a 2-D grid  $2048 \times  1024$ (unless stated otherwise).
 Protons are advances using
the Boris' scheme with a time step $\Delta t=0.05 /\omega_{c\mathrm{p}0}$,
while the magnetic field $\boldsymbol{B}$
is advanced with a smaller time step $\Delta t_B = \Delta t/10$.
The initial ambient magnetic field is in the 2D simulation box,
$\boldsymbol{B}_{0}=B_{0}(\cos\theta_{BR},\sin\theta_{BR},0)$
with the initial angle between the magnetic field and the radial
direction $\theta_{BR}=0^\mathrm{o}$ and $32^\mathrm{o}$. 
Initially we set
  $\beta_{\mathrm{p}\|}=2.4$ and
  $T_{\mathrm{p}\perp}/T_{\mathrm{p}\|}= 0.7$ to start the simulation close
to the unstable region and we set the expansion time $t_e=10^4 /\omega_{c\mathrm{p}0}$
(unless stated otherwise).

\subsection{Radial magnetic field}
\label{radial}
We start with the well-studied case with the strictly radial magnetic field \citep{mattal06,hetr08}.
In this case the expansion continuously drives $T_{\mathrm{p}\|}>T_{\mathrm{p}\perp}$.
When no wave activity/turbulence is present, the system is collisionless,
and no important heat flux exists one expects that the plasma system
follows the double adiabatic/CGL
prediction \citep{chewal56} for the parallel and perpendicular proton temperatures 
\begin{equation}
\left.T_{\mathrm{p}\|}\right|_{CGL} \propto \frac{n^2}{B^2} \ \ \mathrm{and} \ \
\left.T_{\mathrm{p}\perp}\right|_{CGL}\propto B, \label{CGL}
\end{equation}
respectively.
For the radial magnetic field we have $B\propto R^{-2}$ and $n\propto R^{-2}$
so that $T_{\mathrm{p}\|}$ is constant and $T_{\mathrm{p}\perp}\propto R^{-2}$.

\begin{figure}%[htb]
\centerline{\includegraphics[width=10cm]{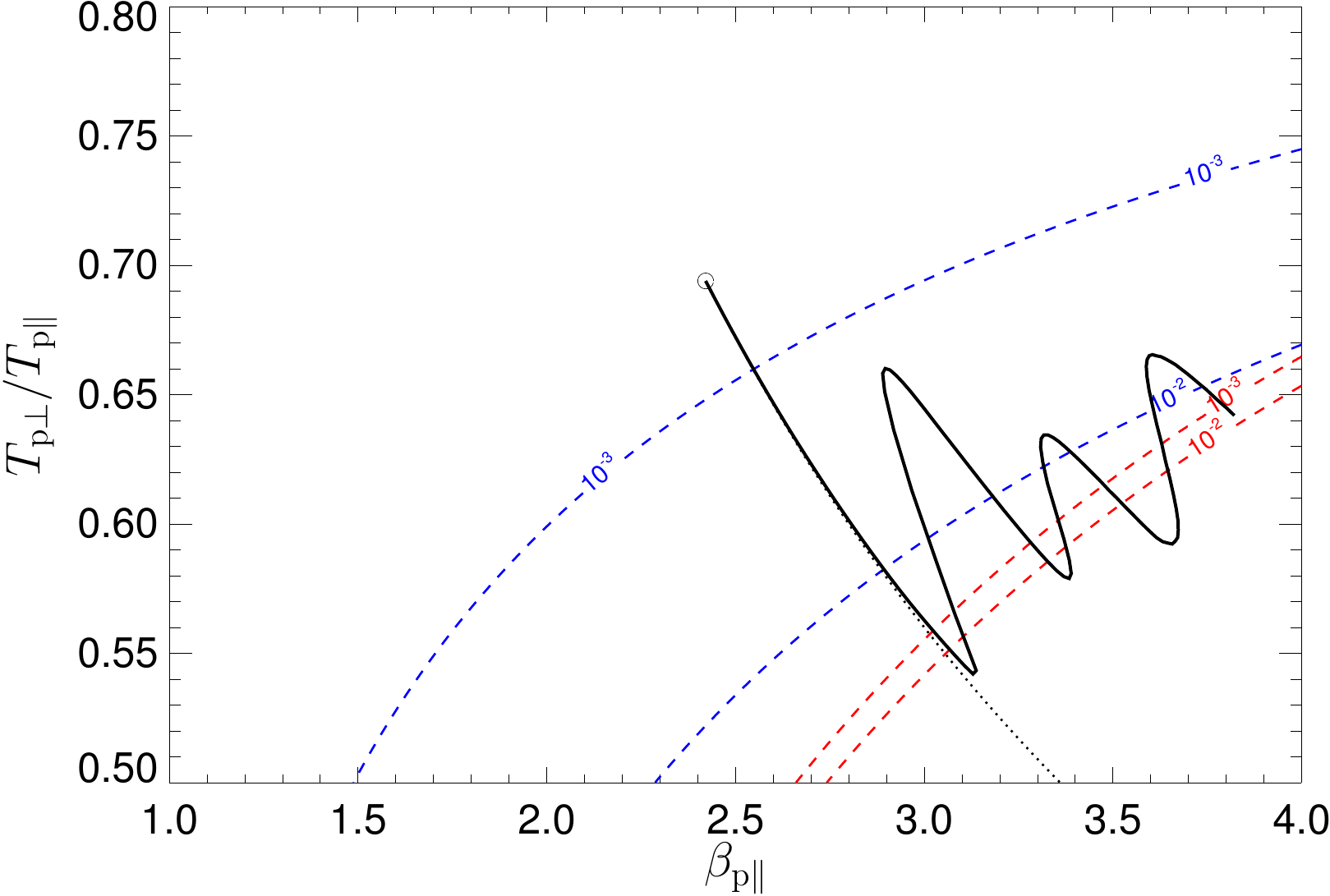}}
\caption{
Evolution in 2-D HEB simulation for the radial $\boldsymbol{B}$ and $t_e \omega_{c\mathrm{p}0}=10^4$: Path
in the space  $(\beta_{\mathrm{p}\|},T_{\mathrm{p}\perp}/T_{\mathrm{p}\|})$
is shown by the solid curve; the empty circle denotes the initial condition.
The dashed contours shows the linear prediction in a homogeneous plasma with
bi-Maxwellian protons, the maximum growth rate (in units of
$\omega_{c\mathrm{p}}$) as
a function of $\beta_{\mathrm{p}\|}$ and $T_{\mathrm{p}\perp}/T_{\mathrm{p}\|}$
for (blue) the parallel proton fire hose and (red) the oblique one.
The dotted curve displays the corresponding double-adiabatic prediction.
\label{anbepar}
}
\end{figure}

Figure~\ref{anbepar} displays the evolution in 2-D HEB simulation for the radial magnetic field ($\theta_{BR}=0^\mathrm{o}$):
the solid curve shows the path
in the space  $(\beta_{\mathrm{p}\|},T_{\mathrm{p}\perp}/T_{\mathrm{p}\|})$; the empty circle denotes the initial condition.
The dashed contours shows the linear prediction in a homogeneous plasma with
bi-Maxwellian protons with respect to the (blue) parallel and (red) oblique fire hoses;
the dotted curve displays the corresponding 
double adiabatic/CGL
prediction. The system initially follows the CGL prediction 
$\beta_{\mathrm{p}\|} \propto R^2$, $T_{\mathrm{p}\perp}/T_{\mathrm{p}\|}\propto R^{-2}$.
When the system enters the region unstable with respect to the parallel fire 
hose \citep{garyal98} the generated waves scatter protons and reduce the parallel temperature
which leads to relatively weak departures from the CGL prediction.
The system eventually enters the region unstable with respect to the
oblique fire hose \citep{hema00}. The plasma then jumps towards the stable region and
after that a CGL-like behaviour reappears and the parallel
temperature anisotropy increases. This is stopped by another jump that strongly
reduces the theoretical instability growth rates. The 
oscillations between weakly and strongly unstable repeat 
till the end of the simulations.

\begin{figure}%[htb]
\centerline{\includegraphics[width=11cm]{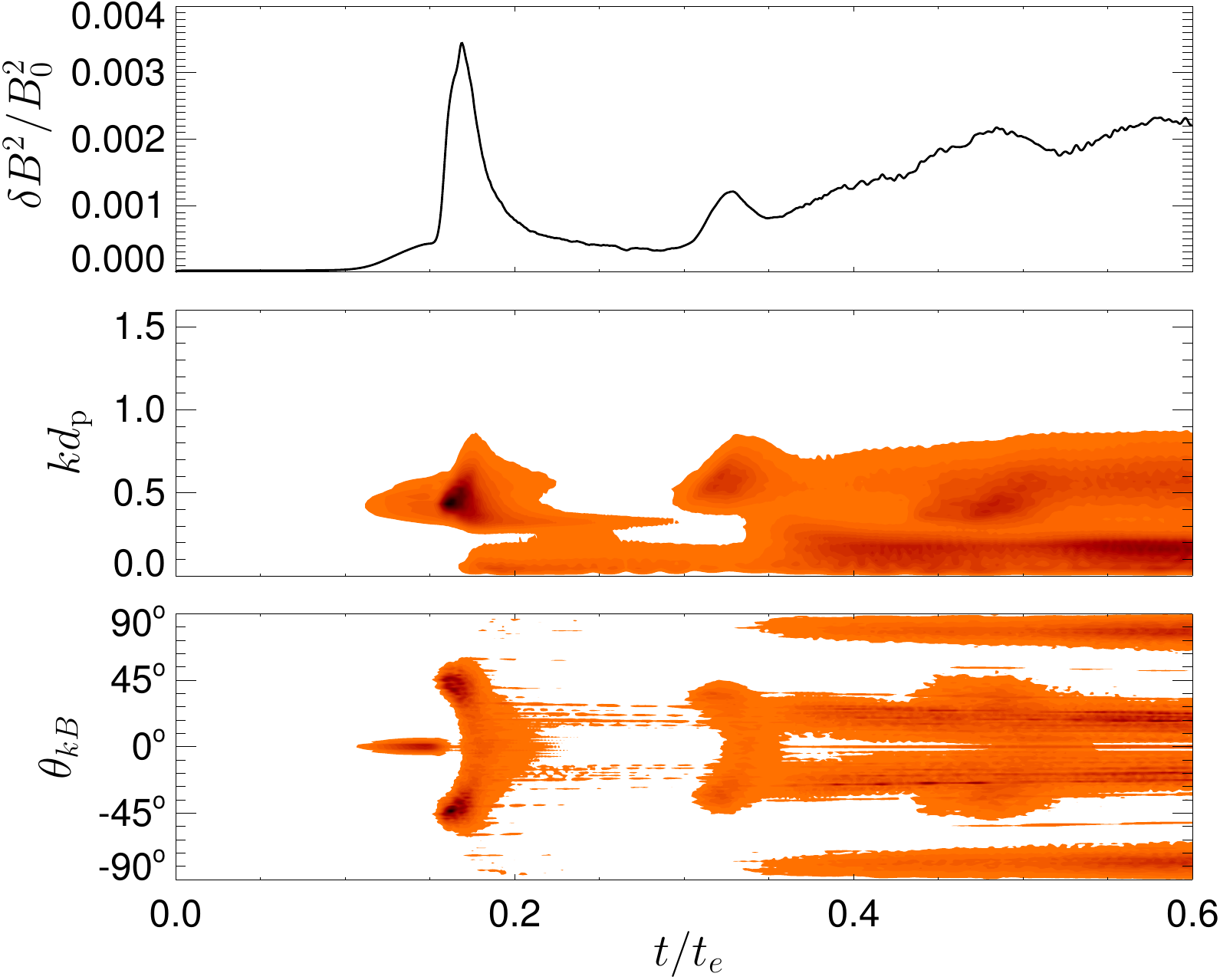}}
\caption{
Evolution in 2-D HEB simulation 
for the radial $\boldsymbol{B}$ and $t_e \omega_{c\mathrm{p}0}=10^4$:
 (top) Fluctuating
magnetic energy $\delta B^2/B_0^2$ as a function of time
(middle)  Color scale plot of the fluctuating magnetic energy
$\delta B^2$ as a function of
time and the wave vector $k$.
(bottom) Color scale plot of the fluctuating magnetic energy
$\delta B^2$ as a function of
time and angle $\theta_{kB}$.
\label{dbpar}
}
\end{figure}

The oscillatory behaviour is connected with the self-destructive properties of the oblique
fire hose.
Figure~\ref{dbpar} shows the evolution of the fluctuating magnetic energy.
The top panel shows the fluctuating
magnetic energy $\delta B^2/B_0^2$ as a function of time.
The middle and bottom panels show
color scale plots of the fluctuating magnetic energy
$\delta B^2$ as a function of
time and the wave vector $k$ and
function of
time and angle $\theta_{kB}$ between the ambient magnetic field and the wave vector, respectively.
As the system expands the plasma becomes unstable with respect to
the parallel fire hose and relatively weak ($\delta B^2/B_0^2\sim 0.0004$)
quasi-parallel ($\theta_{kB}\sim 0^\mathrm{o}$) propagating modes
appear at proton scales ($k\sim 0.5 d_\mathrm{p}^{-1}$) for $0.10\lesssim t/t_e \lesssim 0.15$.
Around $t\sim 0.4$ oblique fire hose wave activity appears and mostly disappears
quite rapidly (on a time scale $\sim 0.1 t_e$). This self-destructive behaviour 
is connected with the dispersive properties of 
the oblique fire hose instability: it destabilizes nonpropagating oblique modes that
only exist for sufficiently anisotropic protons, and, as the generated waves scatter
protons and reduce the temperature anisotropy, the nonpropagating modes disappear
and the fluctuations transforms to damped oblique Alfv\'en waves \citep{hema00,hema01}.
The system with a relatively weak level of wave activity then follows a CGL-like evolution
leading to the generation of the parallel temperature anisotropy and again (mostly oblique) fire hose activity
appears leading to the quasi-periodic oscillations in $\delta B$ and in the space  $(\beta_{\mathrm{p}\|},T_{\mathrm{p}\perp}/T_{\mathrm{p}\|})$.

For the radial magnetic field we recover the previous results
\citep{mattal06,hetr08}.  
The expansion time we used in the HEB simulations is at least 10 times faster
than what is typically observed in the solar wind.
It is important to investigate the role of the expansion time.

\subsection{Role of the expansion time}
\label{radialslow}

We performed another  simulation with $t_e \omega_{c\mathrm{p}0}=10^5$.
This value is close to realistic expansion times in the vicinity of 1 au. 
In order to be able to perform a simulation on such much longer time scales
compared to the previous case we reduced the size of
the simulation box to $512 \times  512$ (while keeping the same resolution
and the same initial plasma parameters).
We also increased the number of particle per cell to 16,384 
 to reduce
the noise level so that the corresponding numerical particle
scattering is weak on the expansion time scale.

\begin{figure}%[htb]
\centerline{\includegraphics[width=10cm]{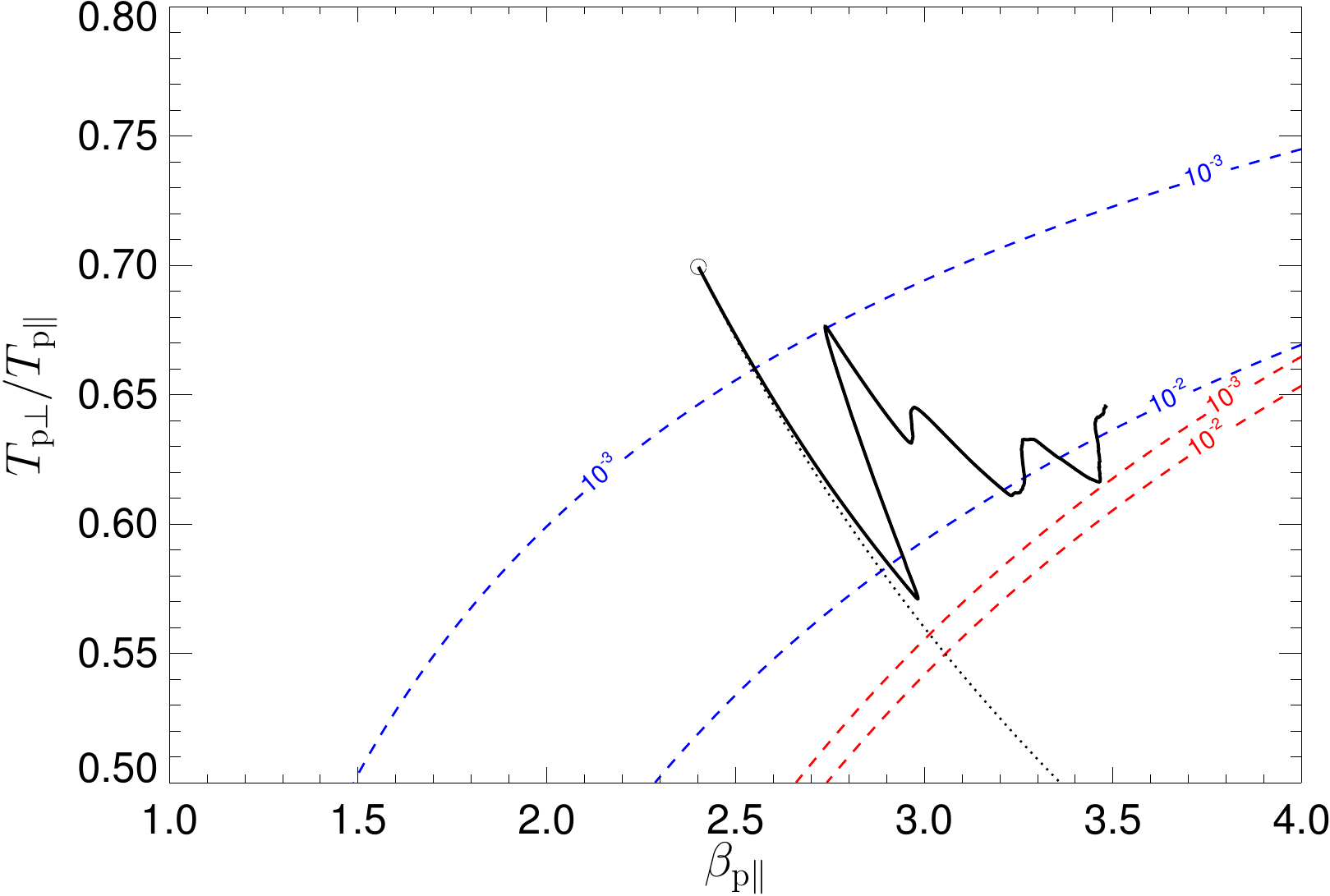}}
\caption{
Evolution in 2-D HEB simulation for the radial $\boldsymbol{B}$ and $t_e \omega_{c\mathrm{p}0}=10^5$: Path
in the space  $(\beta_{\mathrm{p}\|},T_{\mathrm{p}\perp}/T_{\mathrm{p}\|})$
is shown by the solid curve; the empty circle denotes the initial condition.
The dashed contours shows the linear prediction in a homogeneous plasma with
bi-Maxwellian protons, the maximum growth rate (in units of
$\omega_{c\mathrm{p}}$) as
a function of $\beta_{\mathrm{p}\|}$ and $T_{\mathrm{p}\perp}/T_{\mathrm{p}\|}$
for (blue) the parallel proton fire hose and (red) the oblique one.
The dotted curve displays the corresponding double-adiabatic prediction.
\label{anbeparslow}
}
\end{figure}

One expects that for faster expansion the system enters further into the
unstable regions and generates stronger wave activity to counteract
the expansion-driven anisotropization \citep{hetr05,mattal06}.
Figure~\ref{anbeparslow} shows the evolution in 2-D HEB simulation for the radial magnetic field ($\theta_{BR}=0^\mathrm{o}$)
and $t_e \omega_{c\mathrm{p}0}=10^5$ in the same format as in Figure~\ref{anbepar}:
 the solid curve displays the path
in the space  $(\beta_{\mathrm{p}\|},T_{\mathrm{p}\perp}/T_{\mathrm{p}\|})$; the empty circles denote the initial condition.
The dashed contours shows the linear prediction in a homogeneous plasma with
bi-Maxwellian protons for (blue) the parallel and (blue) oblique fire hoses.
The dotted curve displays the corresponding
double adiabatic/CGL
prediction. Figure~\ref{anbeparslow} indeed demonstrates that the system
with the slower expansion enters less far into the unstable region. 
Moreover, this plot suggests that the system does not enter inside the
region unstable with respect to the oblique fire hose. On the
other hand, the observed oscillations suggest that the oblique fire hose
was active.

\begin{figure}%[htb]
\centerline{\includegraphics[width=11cm]{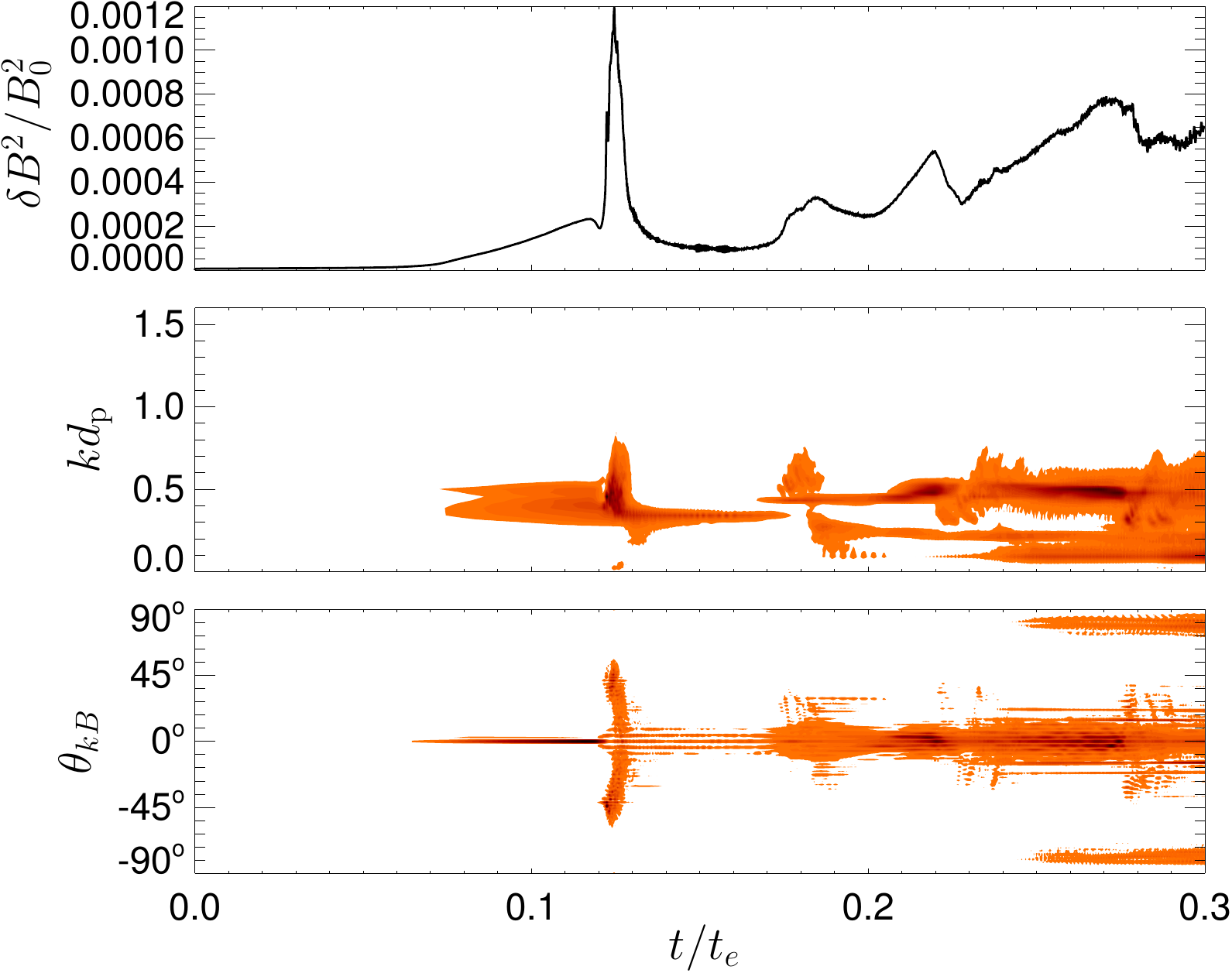}}
\caption{
Evolution in 2-D HEB simulation with the radial magnetic field and $t_e \omega_{c\mathrm{p}0}=10^5$:
 (top) Fluctuating
magnetic energy $\delta B^2/B_0^2$ as a function of time
(middle)  Color scale plot of the fluctuating magnetic energy
$\delta B^2$ as a function of
time and the wave vector $k$.
(bottom) Color scale plot of the fluctuating magnetic energy
$\delta B^2$ as a function of
time and angle $\theta_{kB}$.
\label{dbparslow}
}
\end{figure}

Figure~\ref{dbparslow} shows the evolution of the fluctuating magnetic energy in
the same format as in Figure~\ref{dbpar}.
The top panel shows the fluctuating
magnetic energy $\delta B^2/B_0^2$ as a function of time.
The middle and bottom panels show
color scale plots of the fluctuating magnetic energy
$\delta B^2$ as a function of
time and the wave vector $k d_\mathrm{p}$ and as
a function of
time and angle $\theta_{kB}$ between the ambient magnetic field and the wave vector, respectively.

The evolution is similar to that seen in the simulation with the ten-times faster
expansion (see subsection~\ref{radial}). 
The expansion drives continuously $T_{\mathrm{p}\|}>T_{\mathrm{p}\perp}$ following
the CGL prediction.
When the system enters the region unstable to the parallel fire hose,
the generated wave activity affects the evolution causing small departures
from the CGL prediction. When oblique fire hose activity appears, the
system rapidly moves to the less unstable region while a large portion of generated
wave activity is damped. After that, the system again evolves in a CGL-like 
manner and the process repeats quasi-periodically. 

The comparison between the two simulations demonstrates that the evolution for
$t_e \omega_{c\mathrm{p}0}=10^4$ and 
$t_e \omega_{c\mathrm{p}0}=10^5$ are similar (qualitatively and semiquantitatively).
The main difference is that the amplitude of fluctuations is smaller for
the slower expansion and that the system enters further to the unstable
region for the faster expansion. In the case of $t_e \omega_{c\mathrm{p}0}=10^5$,
the bi-Maxwellian linear theory predicts that the system remains stable with
respect to the oblique fire hose but in the simulation an important
oblique fire hose activity appears. 
We'll look
at this phenomenon in detail in the next subsection, in the
case of the oblique magnetic field.

\subsection{Oblique magnetic field}
\label{oblique}

The magnetic field in the solar wind is usually oblique with
respect to the radial direction. 
For the expanding box approximation
we have from the conservation of the magnetic flux
for radial and transverse component of the magnetic field 
\begin{equation}
B_r \propto R^{-2} \ \ \mathrm{and} \ \ B_t \propto R^{-1},
\end{equation}
respectively, in agreement with the ideal Parker spiral 
model \citep{park58}.
For a magnetic field close to the radial direction 
the expansion drives the parallel anisotropy, $T_{\mathrm{p}\|}>T_{\mathrm{p}\perp}$, but
for a sufficiently oblique magnetic field ($\theta_{BR}\gtrsim 45^\textrm{o}$)
the opposite anisotropy is generated. In this subsection,
we investigate this transition starting with $\theta_{BR}=32^\textrm{o}$.
This particular value of $\theta_{BR}$ corresponds to the radial distance $R\sim0.6$~au and was chosen 
in order to have initially a sufficiently long time of the generation of $T_{\mathrm{p}\|}>T_{\mathrm{p}\perp}$
and a relatively long phase where $T_{\mathrm{p}\|}<T_{\mathrm{p}\perp}$ is generated at later times
of the simulations.
The duration of this simulation is $t_{max}=2 t_e$ so that
at the end $R\sim 1.8$~au and the ambient magnetic field angle with the radial
direction is $\theta_{BR}\sim 62^\textrm{o}$.

\begin{figure}%[htb]
\centerline{\includegraphics[width=10cm]{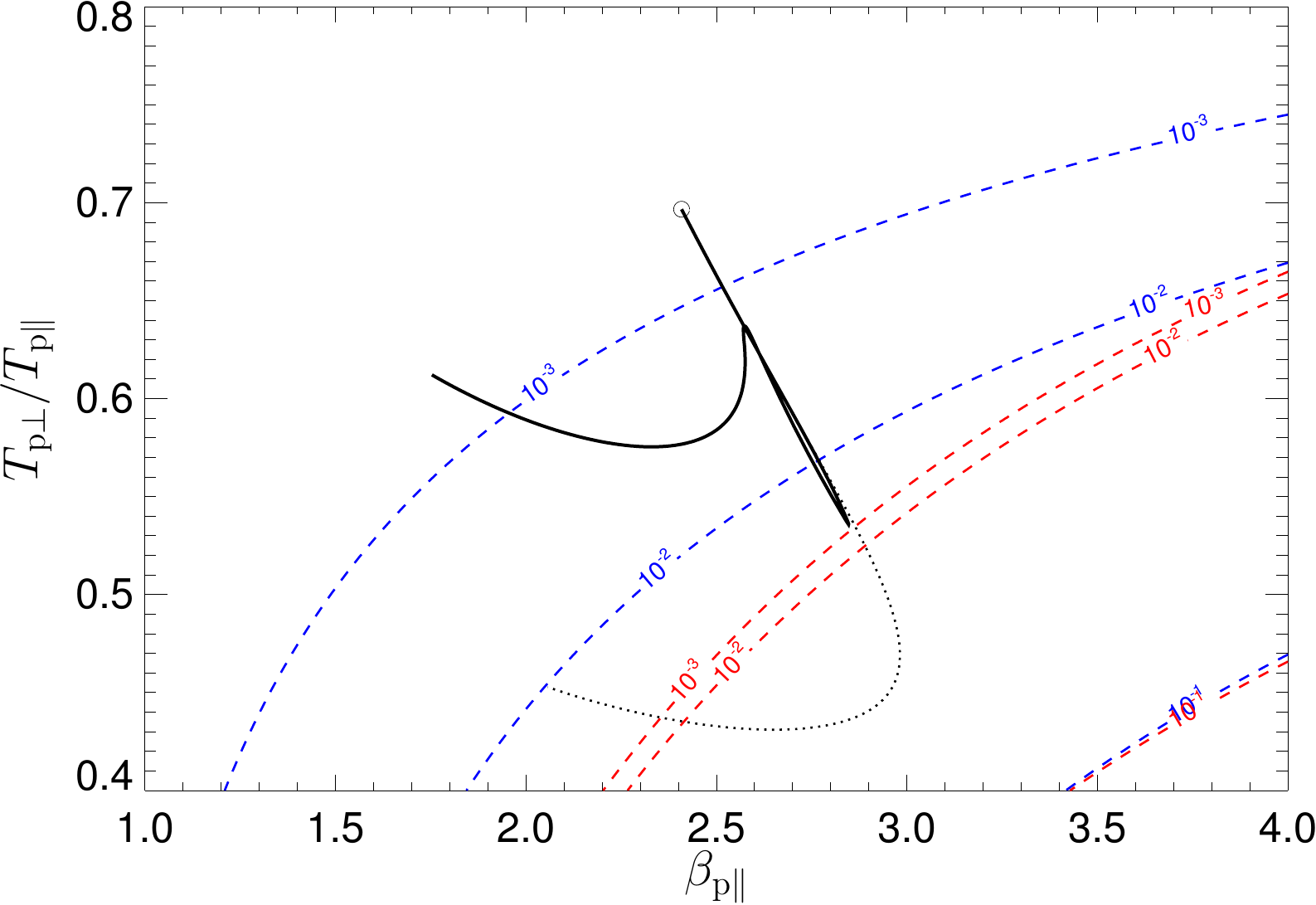}}
\caption{
Evolution in 2-D HEB simulation for  $\theta_{BR}=32^\textrm{o}$ and $t_e \omega_{c\mathrm{p}0}=10^4$: Path
in the space  $(\beta_{\mathrm{p}\|},T_{\mathrm{p}\perp}/T_{\mathrm{p}\|})$
is shown by the solid curve; the empty circle denotes the initial condition.
The dashed contours shows the linear prediction in a homogeneous plasma with
bi-Maxwellian protons, the maximum growth rate (in units of
$\omega_{c\mathrm{p}}$) as
a function of $\beta_{\mathrm{p}\|}$ and $T_{\mathrm{p}\perp}/T_{\mathrm{p}\|}$
for (blue) the parallel proton fire hose and (red) the oblique one.
The dotted curve displays the corresponding double-adiabatic prediction.
\label{anbe}
}
\end{figure}

Figure~\ref{anbe} shows the evolution in 2-D HEB simulation with $\theta_{BR}=32^\textrm{o}$
in the same format as in Figures~\ref{anbepar} and~\ref{anbeparslow}:
 the solid curve displays  the path
in the space  $(\beta_{\mathrm{p}\|},T_{\mathrm{p}\perp}/T_{\mathrm{p}\|})$; 
the empty circle denotes the initial condition.
The dashed contours shows the linear prediction in a homogeneous plasma with
bi-Maxwellian protons, the maximum growth rate (in units of
$\omega_{c\mathrm{p}}$) as
a function of $\beta_{\mathrm{p}\|}$ and $T_{\mathrm{p}\perp}/T_{\mathrm{p}\|}$
for (blue) the parallel proton fire hose and (red) the oblique one.
The dotted curve displays the double-adiabatic prediction.
The system initially follows the double-adiabatic prediction but
when it enters the region unstable with respect to the parallel
fire hose this behaviour is slightly affected. The presence
of this instability is not sufficient to overcome the temperature
anisotropization owing to the expansion in this case. The system enters less far
into the unstable region compared to the radial case with the same expansion
time (see Figure~\ref{anbepar}) as the anisotropization in the case of
the oblique magnetic field is weaker. The plasma does not
enter into the region unstable with respect to the oblique fire hose
in the corresponding bi-Maxwellian plasma. However, the path
in $(\beta_{\mathrm{p}\|},T_{\mathrm{p}\perp}/T_{\mathrm{p}\|})$
exhibit a return movement along almost the same CGL trajectory to the less unstable region,
likely due to the destabilization of the oblique fire hose as
in the previous subsection. After this, the system follows a double-adiabatic-like
path, just shifted in the $(\beta_{\mathrm{p}\|},T_{\mathrm{p}\perp}/T_{\mathrm{p}\|})$
space.

\begin{figure}%[htb]
\centerline{\includegraphics[width=10cm]{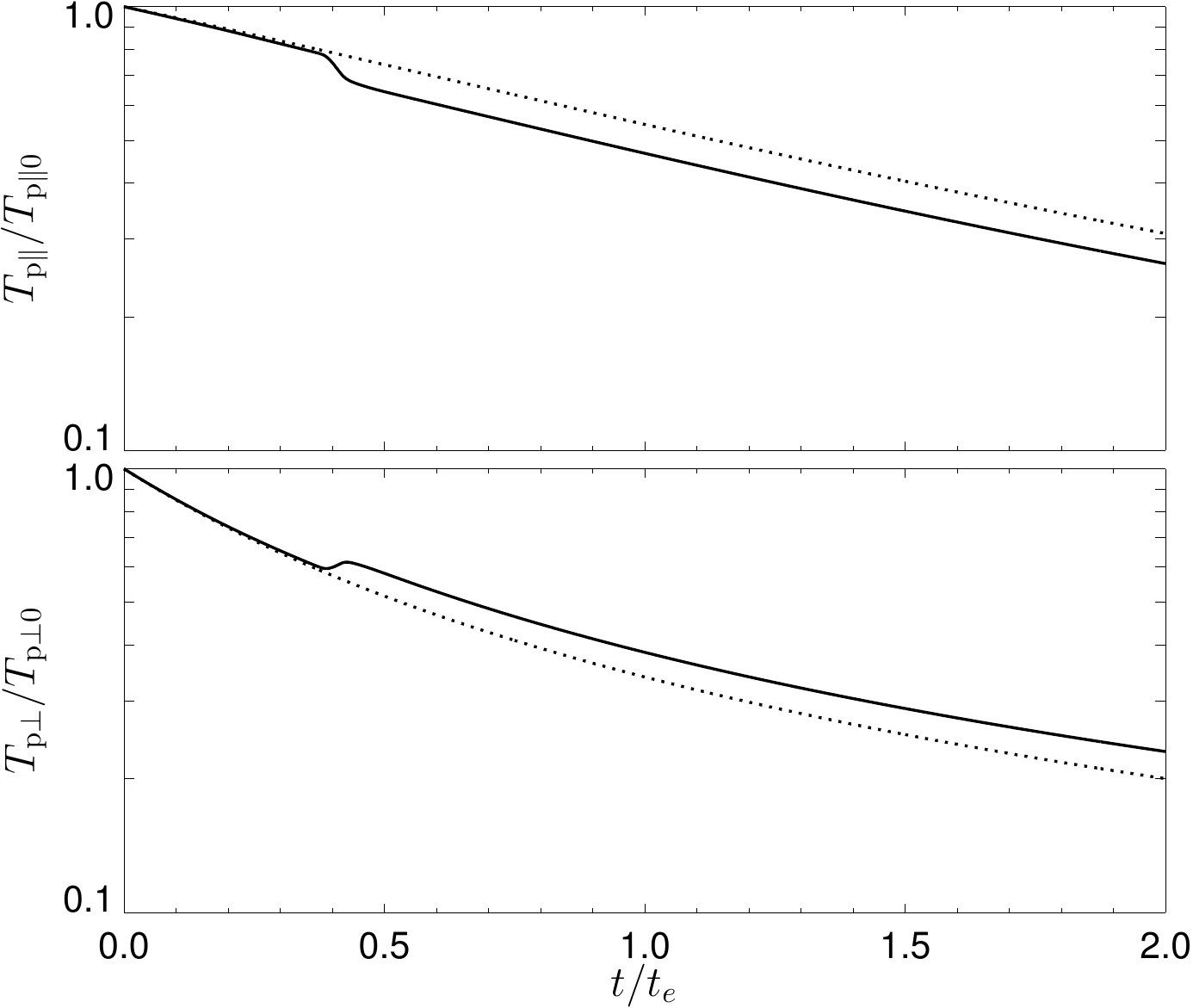}}
\caption{Evolution in 2-D HEB simulation for  $\theta_{BR}=32^\textrm{o}$ and $t_e \omega_{c\mathrm{p}0}=10^4$:
The proton parallel $T_{\mathrm{p}\|}$ (top) and perpendicular $T_{\mathrm{p}\perp}$ (bottom)
as functions of time. 
The dotted curves show the corresponding double-adiabatic predictions.
\label{temp}
}
\end{figure}

A more detailed comparison between the simulation results and the double-adiabatic prediction
is shown in Figure~\ref{temp}
that displays the proton parallel $T_{\mathrm{p}\|}$ (top) and perpendicular $T_{\mathrm{p}\perp}$ (bottom)
as functions of time compared to the prediction.
Until about $t=0.4 t_e$, both the temperatures follow quite closely the double-adiabatic prediction,
while around this time there is an important reduction of the parallel temperature and enhancement of
the perpendicular one compared to the double-adiabatic prediction. After that,
the system follows the double-adiabatic prediction again.

\begin{figure}%[htb]
\centerline{\includegraphics[width=11cm]{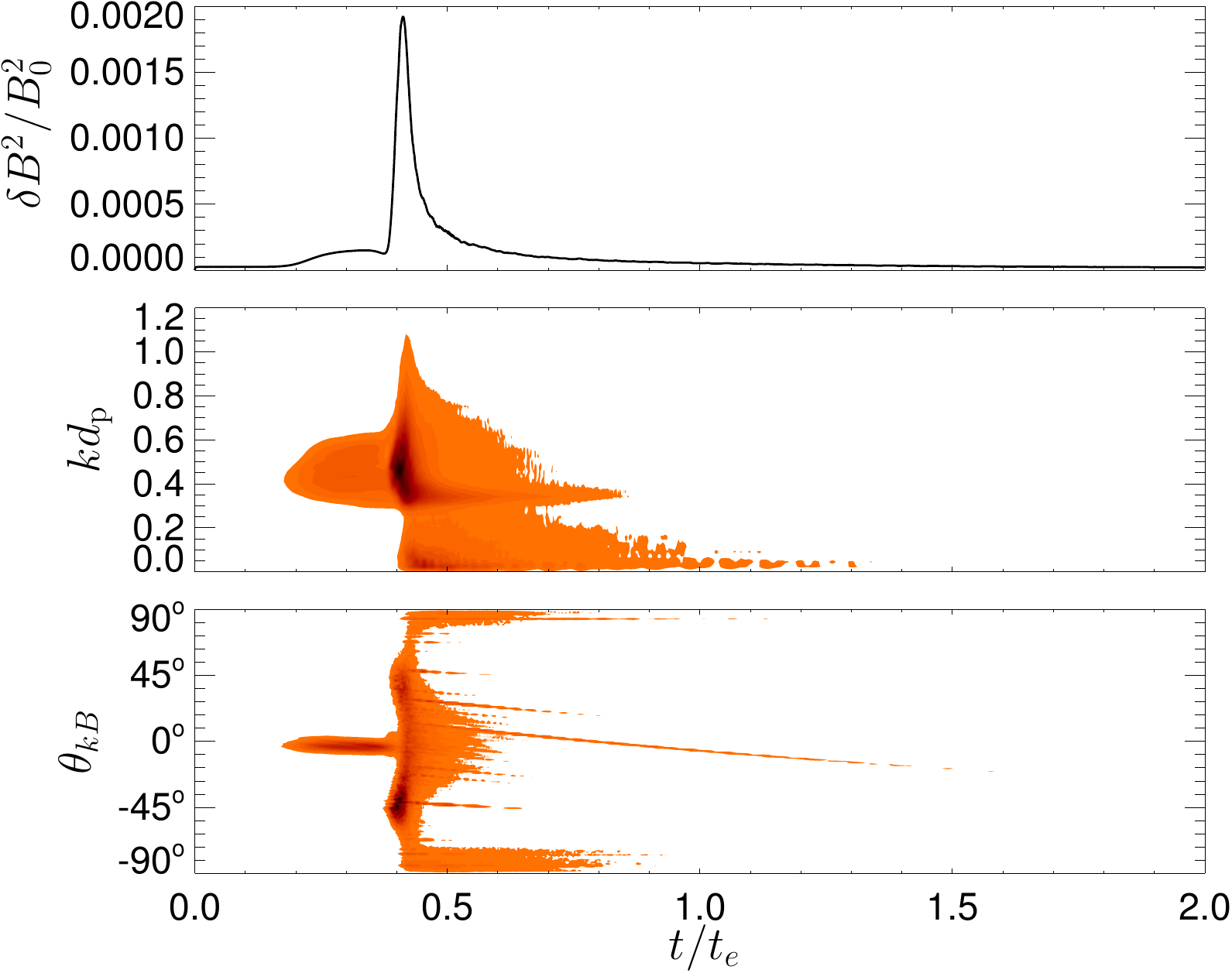}}
\caption{
Evolution in 2-D HEB simulation for  $\theta_{BR}=32^\textrm{o}$ and $t_e \omega_{c\mathrm{p}0}=10^4$:
 (top) Fluctuating
magnetic energy $\delta B^2/B_0^2$ as a function of time
(middle)  Color scale plot of the fluctuating magnetic energy
$\delta B^2$ as a function of
time and the wave vector $k$.
(bottom) Color scale plot of the fluctuating magnetic energy
$\delta B^2$ as a function of
time and angle $\theta_{kB}$.
\label{db}
}
\end{figure}

The disruption of the double-adiabatic prediction is caused by
the generation of the oblique fire hose fluctuations.
Figure~\ref{db} shows the evolution of the fluctuating magnetic energy.
The top panel shows the fluctuating
magnetic energy $\delta B^2/B_0^2$ as a function of time.
The middle and bottom panels show
color scale plots of the fluctuating magnetic energy
$\delta B^2$ as a function of
time and the wave vector $k d_\mathrm{p}$ and as a
function of
time and angle $\theta_{kB}$ between the ambient magnetic field and the wave vector, respectively.
As the system expands, the plasma becomes unstable with respect to
the parallel fire hose and relatively weak ($\delta B/B_0\sim 0.01$) 
quasi-parallel ($\theta_{kB}\sim 0^\mathrm{o}$) propagating modes
appear at proton scales ($k\sim 0.5 d_\mathrm{p}^{-1}$) for $0.2\lesssim t/t_e \lesssim 0.4$.
Around $t\sim 0.4$, oblique fire hose wave activity appears and mostly disappears
quite rapidly (on a time scale $\sim 0.1 t_e$), similarly to the case
with the radial magnetic field.

\begin{figure}%[htb]
\centerline{\includegraphics[width=11cm]{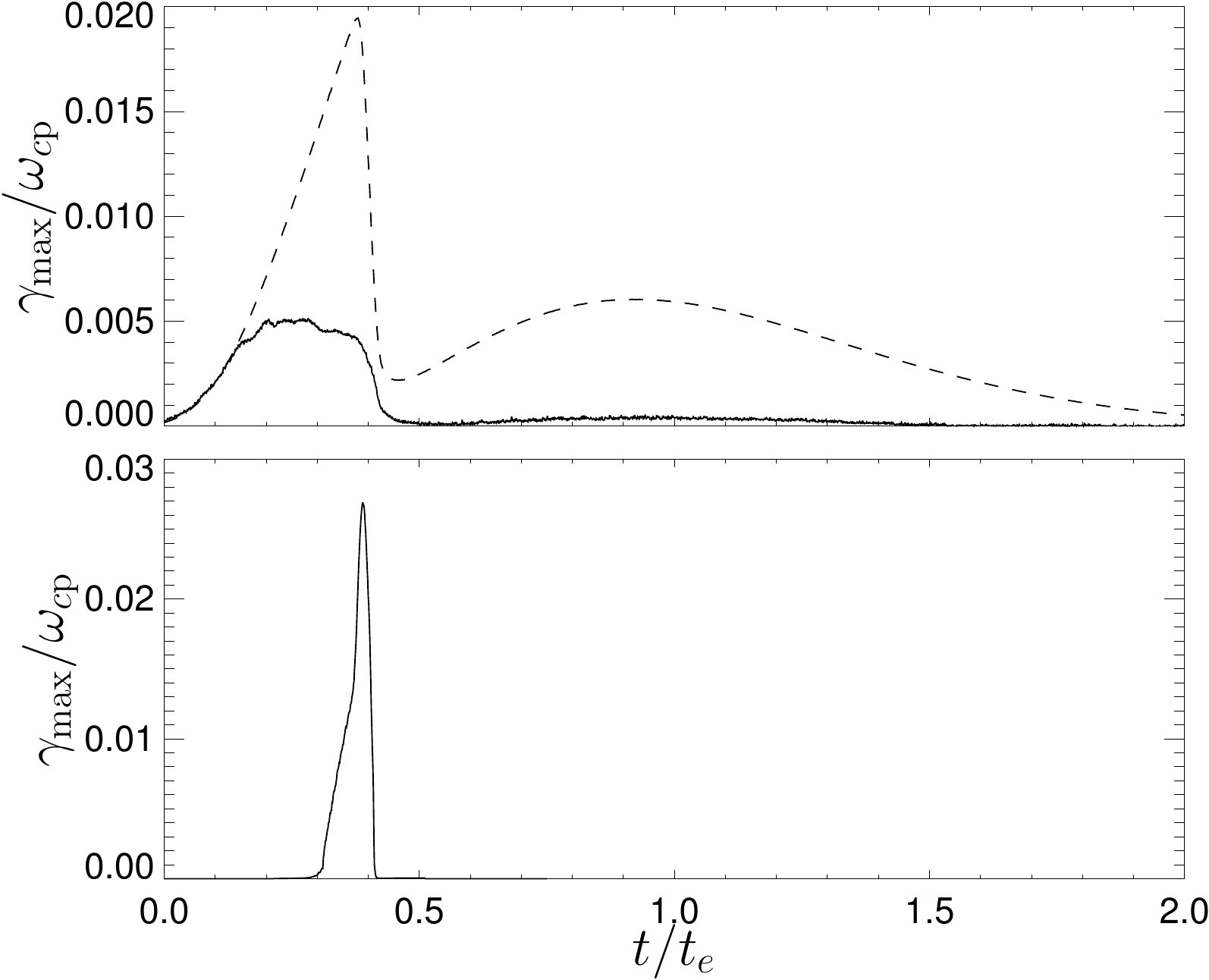}}
\caption{Evolution in 2-D HEB simulation for  $\theta_{BR}=32^\textrm{o}$ and $t_e \omega_{c\mathrm{p}0}=10^4$:
The maximum growth rate $\gamma_\mathrm{max}$ as a function
of time for 
the parallel (top panel) and oblique (bottom panel) fire hose instability.
The solid lines display the linear prediction calculated
from the proton velocity distribution function whereas
the dashed line shows the results for a 
 bi-Maxwellian velocity distribution function corresponding to the parallel
and perpendicular pressures.
\label{gamma}
} 
\end{figure}

Figure~\ref{anbe} reveals a discrepancy between
the linear prediction based on the bi-Maxwellian approximation and the simulation results. As in the case of slower expansion
(subsection~\ref{radialslow}), the linear theory based on the bi-Maxwellian proton velocity distribution
function predicts that the system is all the time stable with respect to 
the oblique fire hose, whereas the simulation results clearly exhibit
a wave activity driven by this instability. 
This is connected with the resonant character of the instability \citep{gary93}; its growth
rate depends on the details (gradients) of the particle velocity
distribution function.
Figure~\ref{gamma} shows the 
maximum growth rate $\gamma_\mathrm{max}$ as a function
of time for
the parallel (top panel) and oblique (bottom panel) fire hose instability.
The solid lines display the linear prediction calculated
from the distribution function observed in the code \cite[cf.,][]{hetr11}, whereas
the dashed line shows the results for 
 bi-Maxwellian velocity distribution functions that have the observed parallel
and perpendicular proton temperatures.

Figure~\ref{gamma} clearly demonstrates that the growth rates calculated from the velocity
distribution functions can widely differ from those calculated from bi-Maxwellian
distribution functions having the same temperatures \cite[cf.][]{isen12}. For the parallel fire hose the two growth rates
are the same only initially, $t\lesssim 0.15 t_e$. As the parallel fire hose activity
increases, protons are scattered mainly in the resonant region of 
the velocity distribution function, giving rise to a quasi-linear plateau-like
distribution. This plateau does not strongly influence the macroscopic
moments so that the linear prediction based on the bi-Maxwellian shape
gives much stronger growth rates. On the other hand, the linear
theory based on the instantaneous velocity distribution predicts an important
growth rate for the oblique fire hose instability for $0.3 t_e \lesssim t \lesssim 0.4 t_e$,
whereas in the corresponding bi-Maxwellian case this mode is stable. 
This behaviour is caused by the interplay between the expansion and the fire hose
instabilities. The expansion affects the whole distribution function whereas 
the two fire hose instabilities are resonant and interact preferably with 
the particles from their resonant regions which are different; the parallel fire hose
interacts through the anomalous cyclotron resonance whereas the oblique fire hose
interacts through the cyclotron resonance \citep[and the Landau resonance after the branch
change, cf.,][]{hetr06}. 
As the parallel fire hose is stabilized through the modification of the distribution 
function in the vicinity of its resonant region, the rest of the distribution function
is almost unaffected and may destabilize the oblique fire hose. At the same time, the parallel fire hose
reduces the anisotropy on the macroscopic level so that in the corresponding bi-Maxwellian
plasma the oblique fire hose is stabilized.
The simulation results are in agreement with the linear prediction based
on the instantaneous velocity distribution functions; the parallel fire hose activity
is saturated around $t\sim 0.3 t_e$, the oblique fire hose wave activity
has a maximum amplitude around $t\sim 0.4 t_e$ and after $t\sim 0.4 t_e$
all the wave activity decays. 
 
The calculation of the linear growth rate is based on discrete
velocity distribution functions calculated in the code over the whole simulation box. The
distribution function is evaluated on a $512\times512$ grid in
the velocity space  $(v_\|,v_\perp)$ with respect to the ambient magnetic field.
The maximum parallel and perpendicular velocities (and consequently the velocity grid size) are
determined in the code by the maximum proton velocities. The finite number
of particles per cell and the discretization are sources of different errors.
In the present case, we estimate the error of the growth rate determination from the velocity
distribution function to be of the order  $10^{-4}\omega_{c\mathrm{p}}$,
so that the values in Figure~\ref{gamma} are physically relevant.
 
\begin{figure}%[htb]
\centerline{\includegraphics[width=14cm]{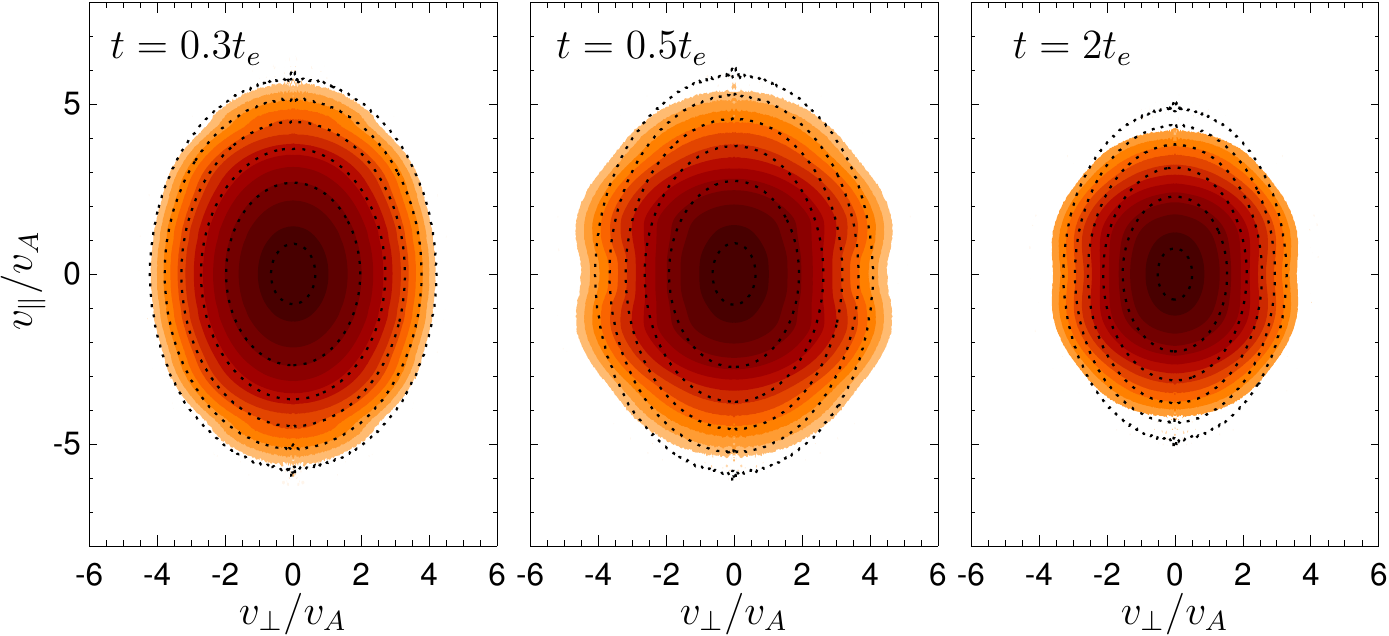}}
\caption{Evolution in 2-D HEB simulation for  $\theta_{BR}=32^\textrm{o}$ and $t_e \omega_{c\mathrm{p}0}=10^4$:
Color scale plots of the proton velocity distribution functions
as functions of  $v_\perp$  and  $v_\|$ (normalized to $v_{A}$)  at
(left)
  $t=0.3 t_e$, (middle)  $t= 0.5 t_e$, and (bottom panels) at $t=2t_e$.
The dotted contours display the double-adiabatic prediction.
\label{df}
} 
\end{figure}

The actual shape of the proton velocity distribution function is shown in
Figure~\ref{df}, that displays color scale plots of the distribution functions
as functions of  $v_\perp$  and  $v_\|$ (normalized to the Alfv\'en velocity $v_{A}$)  at
(left)
  $t=0.3 t_e$, (middle)  $t= 0.5 t_e$, and (bottom panels) at $t=2t_e$.
The dotted contours display the double-adiabatic prediction for comparison
(this plot was obtained from the velocity distribution function at $t=0$ and
rescaled using Equation (\ref{CGL})).
At $t = 0.3 t_e$ there are small (hardly discernible) differences between
the distribution function and the CGL prediction around $|v_\||\sim 0.4 v_A$.
At $t = 0.5 t_e$, when most of the wave activity driven by the two fire hose
is damped, there are clear differences between the distribution
function and the CGL. Afterwards, the distribution function essentially
only rescales, following the CGL prediction. 

\section{Discussion}
\label{discussion}

We presented three two-dimensional hybrid expanding box simulations
of the competition between the expansion-driven parallel proton temperature 
anisotropy and fire hose instabilities. We used similar plasma parameters and
varied the expansion timescale and the (initial) angle between the ambient
magnetic field and the radial direction. The simulation results of the three
simulations have similar properties; as the expansion drives $T_{\mathrm{p}\|}>T_{\mathrm{p}\perp}$,
the system becomes unstable with respect the dominant parallel fire hose
instability. This instability is generally unable to counteract the induced
anisotropization and the system becomes eventually unstable with respect 
to the oblique fire hose. This instability efficiently reduces
the anisotropy and the system becomes stable, while a significant part of the generated
electromagnetic fluctuations is damped to protons. As long as the magnetic
field is sufficiently close to the radial direction, this evolution repeats itself
and the electromagnetic fluctuations accumulates. For sufficiently oblique
magnetic field the expansion drives $T_{\mathrm{p}\perp}>T_{\mathrm{p}\|}$ and brings the system to
the stable region with respect to the fire hose instabilities, where
the generated wave activity is damped. This evolution may eventually
lead to instabilities driven by $T_{\mathrm{p}\perp}>T_{\mathrm{p}\|}$ \citep{traval07b}. 
For slower driver (longer expansion time) and/or oblique magnetic field, the system
enters less far into the unstable region and generates weaker wave amplitudes. Starting
with a sufficiently oblique magnetic field, the system may completely avoid the fire hose instabilities. 
The simulation results also indicate that assumption of bi-Maxwellian velocity distribution
functions may lead to wrong results, as both the fire hoses are resonant and the simulation results show that the linear
 growth rate is sensitive to the details of the proton distribution function \cite[cf.][]{hetr11}.
We expect similar results for alpha particle driven fire hose instabilities
\citep{marual12,versal13,mattal15b}. 
A differential velocity between different ion species
is another source of free energy for instabilities \citep{mattal13}.
For electromagnetic instabilities the important parameter
is the ratio between the differential and Alfv\'en velocities 
\citep{gary93,daga98}.
Similar to the case of the parallel temperature anisotropy
the double adiabatic approximation predicts that this ratio increases with the radial
distance for the magnetic field that is not too oblique with respect
to the magnetic field
\citep{mattal12}. For sufficiently oblique magnetic field this
ratio decreases owing to the rotational force 
\citep{versal15}.

The present simulation results are obtained in a collisionless, homogeneous plasma where 
Coulomb collisions and the presence
of an important turbulent solar wind activity \citep{brca13} are neglected.
We note that  a behaviour similar to that presented here
is also seen in weakly collisional plasmas \citep{hetr15} and
simulations results of \cite{hellal15} indicate that kinetic instabilities,
such as the oblique fire hose, coexist with fully developed turbulence. 
Moreover, similar results are also obtained for a completely different driver (velocity shear)
of the temperature anisotropy \citep{kunzal14}.
Consequently,
our results are quite robust and applicable to a wide range of space or astrophysical plasmas. 

In the context of the present work, it is interesting to briefly discuss the results of \cite{mattal07}. 
The observed change of proton behaviour around 1~au in the $(\beta_{\mathrm{p}\|},T_{\mathrm{p}\perp}/T_{\mathrm{p}\|})$ plane 
is likely partly connected with the change of 
the expansion-driven anisotropization around $\theta_{BR}\sim 45^\mathrm{o}$ that corresponds roughly to 1~au.
However, these observations also indicate that $\beta_{\mathrm{p}\|}$ increases with
the radial distance even beyond 1~au whereas the double adiabatic approximation predicts
that it should decrease then. This is probably connected with the fact that the solar wind protons do not
generally follow the double adiabatic prediction since they are continuously heated,
likely owing to the turbulent cascade.
More work is needed to understand the behaviour of turbulence in the expanding solar wind
with an oblique magnetic field, the resulting proton energization, and the role
of the fire hose instabilities in such a complex system.

\section*{Acknowledgements}
The author acknowledges grant 15-10057S of the Grant Agency of the Czech Republic
and projects RVO:67985815 and RVO:68378289.

\end{document}